\pdfoutput=1
\documentclass[sigconf,authorversion,nonacm]{acmart} 

\AtBeginDocument{%
  \providecommand\BibTeX{{%
    \normalfont B\kern-0.5em{\scshape i\kern-0.25em b}\kern-0.8em\TeX}}}

\usepackage{listings} 
\usepackage{caption}
\usepackage{subcaption}
\usepackage{algorithm}
\usepackage{algpseudocode}
\usepackage[htt]{hyphenat}

\begin{document}

\title{Large Scale Generation of Labeled Type Data for Python}

\author{Ibrahim Abdelaziz}
\affiliation{%
 \institution{IBM TJ Watson Research Center}
 \city{Yorktown Heights, NY}
 \country{United States}
}
\email{ibrahim.abdelaziz1@ibm.com}

\author{Julian Dolby}
\affiliation{%
 \institution{IBM TJ Watson Research Center}
 \city{Yorktown Heights, NY}
 \country{United States}
}
\email{dolby@us.ibm.com}

\author{Kavitha Srinivas}
\affiliation{%
 \institution{IBM TJ Watson Research Center}
 \city{Yorktown Heights, NY}
 \country{United States}
}
\email{kavitha.srinivas@ibm.com}

\renewcommand{\shortauthors}{Trovato and Tobin, et al.}

\begin{abstract}
Recently, dynamically typed languages, such as Python, have gained unprecedented popularity. Although these languages alleviate the need for mandatory type annotations, types still play a critical role in program understanding and preventing runtime errors.  An attractive option is to infer types automatically to get static guarantees without writing types.  Existing inference techniques rely mostly on static typing tools such as \texttt{PyType} for direct type inference; more recently, neural type inference has been proposed.  However, neural type inference is data hungry, and depends on collecting labeled data based on static typing. Such tools, however, are poor at inferring user defined types. Furthermore, type annotation by developers in these languages is quite sparse. In this work, we propose novel techniques for generating high quality types using 1) information retrieval techniques that work on well documented libraries to extract types and 2) usage patterns by analyzing a large repository of programs. Our results show that these techniques are more precise and address the weaknesses of static tools, and can be useful for generating a large labeled dataset for type inference by machine learning methods. F1 scores are 0.52-0.58 for our techniques, compared to static typing tools which are at 0.06, and we use them to generate over 37,000 types for over 700 modules.
\end{abstract}



\maketitle

\section{Introduction}

\begin{figure}[htb]
\includegraphics[scale=0.8]{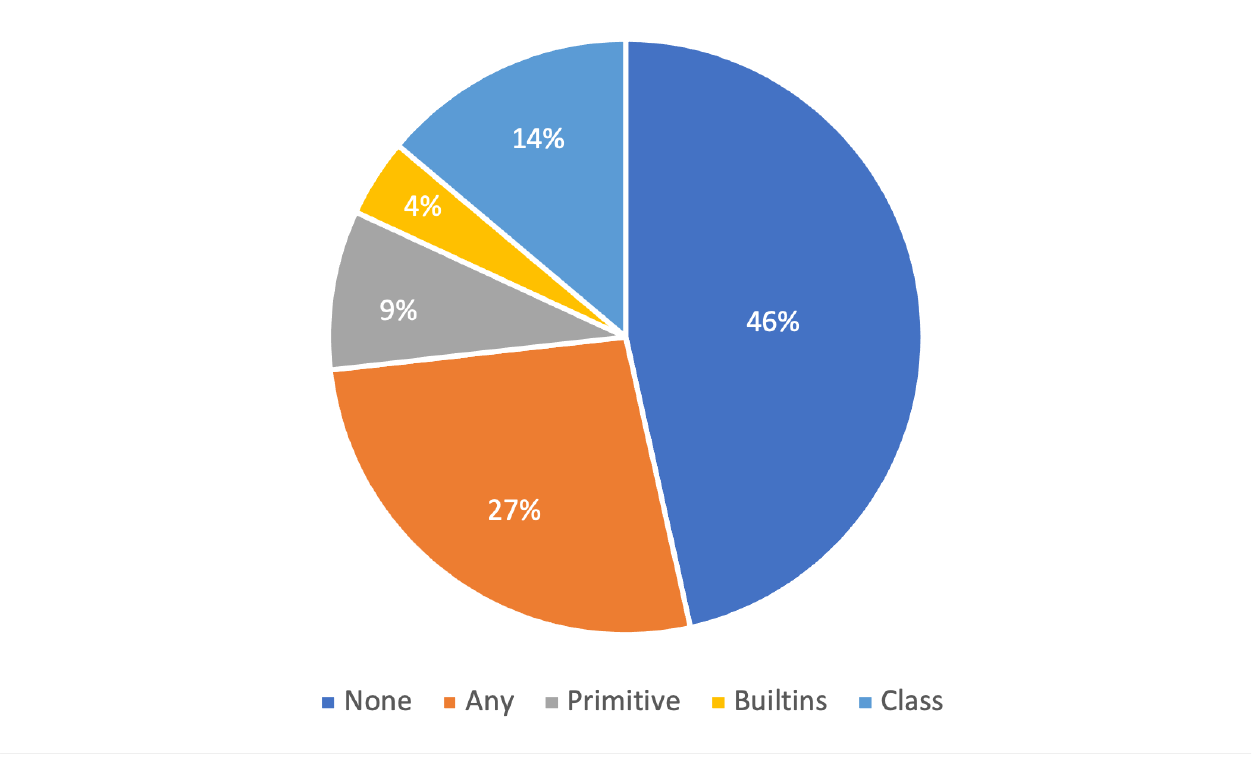}
\caption{PyType predictions for a dataset of 408 repositories}
\label{fig:pytype}

\end{figure}

Dynamically typed languages such as Python have become very popular\footnote{https://stackoverflow.blog/2017/09/06/incredible-growth-python/}, thanks in part to the unprecedented growth of Artificial Intelligence (AI) and the wide adoption of Python for AI frameworks.  
Python, like many dynamic programming languages, does not enforce types statically, but discovers errors only at runtime, which is popular because it allows programmers to build prototypes quickly.  Types, however, are useful for program understanding, for finding errors early and improving program correctness.  Python 3 introduced optional type declarations with PEP484~\cite{PEP484}, but so far there has been little adoption: a recent study showed that fewer than 4\% of repositories from a sample of GitHub had any annotations, and, even in those that did, 80\% of files did not contain a single annotation~\cite{10.1145/3426422.3426981}. Furthermore, traditional type inference has so far proved largely ineffective: the most common tools are \texttt{MyPy} and \texttt{Pytype}, and they infer only 6\% of user provided annotations for types that are not builtins nor primitives. As shown in Figure~\ref{fig:pytype}, they frequently produce \texttt{Any} as a type, which is equivalent to no information.  Furthermore, only 14\% of the types they produce are user defined or library types, which tend to be much more prevalent in user code~\cite{10.1145/3426422.3426981}. 

In this situation, machine learning has become a promising approach; recent systems include Typilus~\cite{typilus} and TypeWriter~\cite{typewriter} perform type inference using neural networks.  However, learning approaches require large amounts of type-annotated code for training, which as we have seen, does not exist.  In fact, neural systems currently rely on tools such as \texttt{Pytype} and \texttt{MyPy}~\cite{typilus} or user specified annotations~\cite{typewriter} for their gold standard.  This labeled data is skewed in ways that will affect the quality of the model that is built, and will provide potentially misleading estimates of accuracy when used as a gold standard.



In this work, we explore techniques to generate high quality types for methods, which can serve as labeled data for data-hungry machine learning approaches. We do believe machine learning techniques have promise for the problem of general type inference.  However, we definitely need better mechanisms to address the problem of obtaining good quality labeled data.  In particular, we explore whether it is possible to (a) extract high quality types from well documented framework code to infer types, and (b) extract high quality types from their usage in open source code.  Our goal here is not so much to propose a mechanism for general type inference as to produce better datasets for use in building better probabilistic type inference systems.

We start with framework data because they are both well used, and well documented.  To infer types from documentation, we use techniques from information retrieval to gather possible types specified in documentation and map them to a set of classes we index. To infer types from usage, we mine usage from millions of programs on GitHub, and explore \textit{duck typing} based on program analysis.  Duck typing refers to the idea that if a class contains \textit{all} the methods called on a given object, then it is a likely candidate for the type of that object.  In other words, if the object walks like a duck and quacks like a duck then it is quite possibly a duck.  Although the idea of duck typing is not new, we apply it in a novel way.  Specifically, we analyze a large repository of 1.3 million code files from GitHub, and combine usage of the same libraries across them.  While we analyze individual programs, we observe how data flows from common API calls to objects returned by the calls \textit{across all programs}.  The duck typing method has the potential to infer user defined types; it cannot infer primitives or builtin types since they do not correspond to classes that we know about.  Type inference from documentation has the potential to do better on builtins and primitives.  Our hypothesis is that by combining these techniques we could infer a greater variety of types, and offset weaknesses in each technique.

We focus on three research questions:
\begin{enumerate}
    \item Do these two techniques yield types that are precise enough to provide high quality labeled data?
    \item Do they yield types that address some of the weaknesses of tools such as \texttt{Pytype} which currently are the state-of-the-art for obtaining labeled type inference data in Python?
    \item Do the two techniques provide non-overlapping sets of types, such that the union of the two approaches increases the size of the labeled set?
\end{enumerate}

Our experiments show that the two techniques produce types for over 37,000 methods and functions in 756 Python modules.  We compare the precision and recall of our type inference techniques against a set of types inferred from dynamic techniques, as well as manual annotations for sample sizes of over 200 functions.  Our F1 scores were .52 and .58 for static analysis and documentation inference, compared to \texttt{PyType} which was .06.  We note that state-of-the-art neural prediction systems such as TypeWriter achieve 0.64 from a trained model based on type annotations.  Our approach is completely unsupervised and we hope that this method of producing labeled data will be helpful for building better neural models for type inference.  The data and code used to create it are publicly available \url{https://github.com/wala/graph4code}.

\section{A Running Example}
\label{sec:example}

\begin{figure}[htb]
\centering



\includegraphics[width=1\columnwidth]{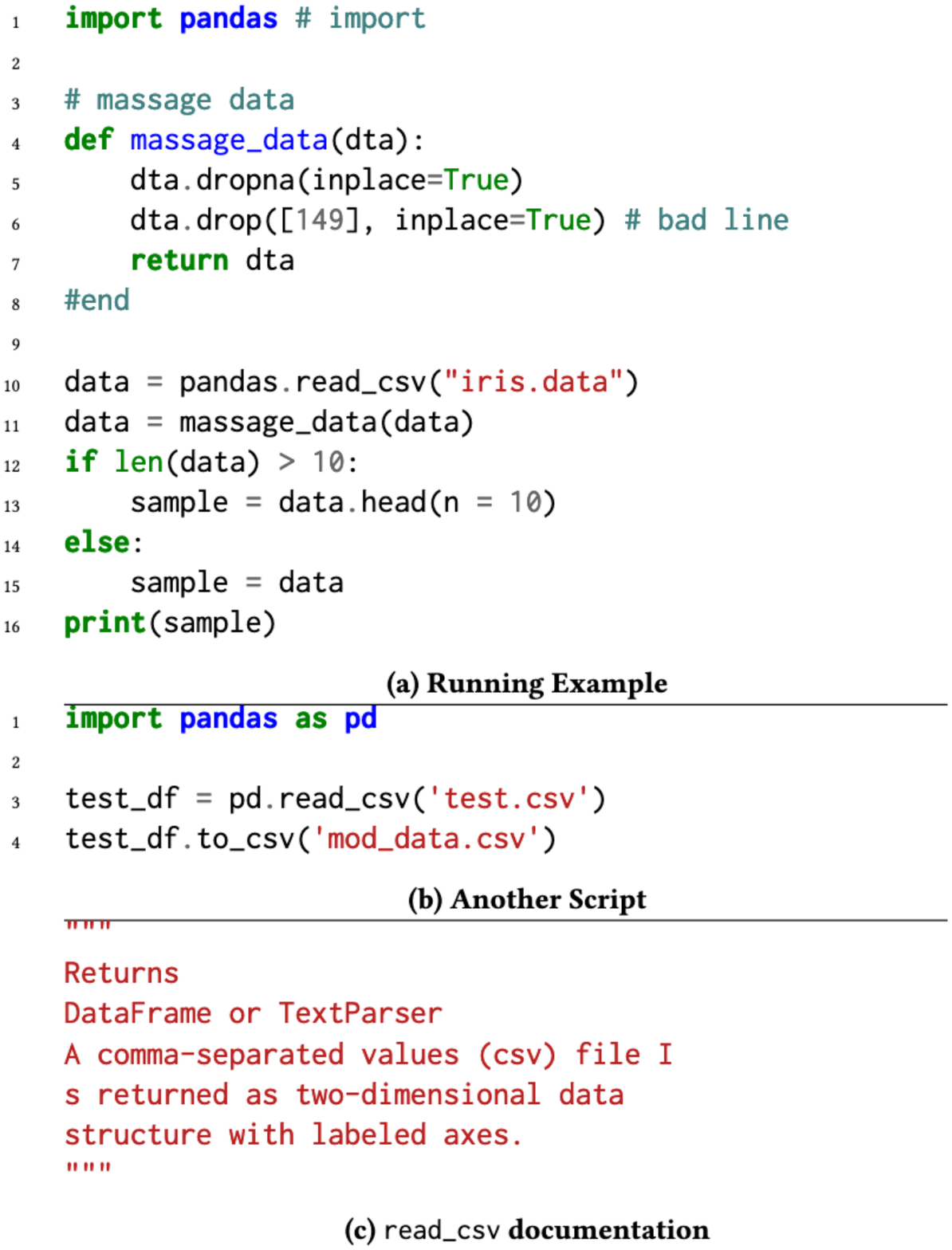}
\caption{Code and documentation example for \texttt{read\_csv}}
\label{running_example}
\end{figure}

Figure~\ref{running_example} shows the core ideas behind large scale generation of labeled types for API calls.  Script 1 in the example calls a function to read a \texttt{pandas.Dataframe} object from the \texttt{pandas} library, and then passes the return value into a function, where the object is used as a receiver for the \texttt{drop} and \texttt{drop\_na} calls.  Script 2 has a more direct relation between the \texttt{read\_csv} call and the \texttt{to\_csv} call on the returned object.  From the perspective of duck typing across multiple scripts, it is clear that the type of returned objects from \texttt{read\_csv} calls must support \texttt{drop}, \texttt{drop\_na}, \texttt{head} and the \texttt{to\_csv}.  From the type definitions of classes of APIs, \texttt{pandas.Dataframe} is clearly a candidate class.  Figure~\ref{running_example} shows the documentation associated with the \texttt{read\_csv} function, and it is clear that the documentation is not formal enough to clearly denote the class being referred to, so we need additional processing to infer the possible type. Classes are mentioned informally, with no reference to their fully qualified name.  They are referred to with natural language using phrases such as \textit{DataFrame or TextParser}, so we need some mechanism to resolve the two classes mentioned here to their fully qualified names (e.g. \texttt{pandas.core.frame.DataFrame}). In our current work, we use simple techniques from information retrieval to find potential types from documentation.

\section{Dataset}
\begin{table*}
    \centering
    \begin{tabular}{l|l}
    \toprule
      Class & Aliases To \\
      \midrule
      statsmodels.datasets.utils.DataFrame   & pandas.core.frame.DataFrame \\
      statsmodels.stats.anova.DataFrame  & pandas.core.frame.DataFrame \\
      bokeh.core.properties.PandasDataFrame & bokeh.core.property.pandas.PandasDataFrame \\
      bokeh.core.properties.PandasDataFrame & bokeh.core.property.pandas.PandasDataFrame \\
      \bottomrule
    \end{tabular}
    \caption{Examples of classes that alias to a different class}
    \label{classes_map}
\end{table*}

Our dataset is based on 1.3 million Python programs on GitHub, which we gathered from Google's public datasets of Python programs from BigQuery with a query to Google BigQuery that focused on Python repositories that more than watch one event in the last year.  The query was issued in August 2019, but reflected a snapshot of GitHub from March 20, 2019, by Google.

To gather relevant classes and methods, we identified the top 500 modules imported in these 1.3 million Python programs.  For each of these modules, we tried to programmatically create a virtual environment, install the module using pip, and then used the python \texttt{inspect} APIs to gather all the classes in the loaded modules, as well as their methods and relevant docstrings.  Python introspect APIs do not just provide classes from the loaded module, they gather classes from the modules that are in the dependency tree of the loaded module.  Furthermore, a quirk of the Python \texttt{inspect} API is that it specifies numerous classes that alias to the same class, based on the dependency of the module. Table~\ref{classes_map} shows such an example - the first two \texttt{DataFrame} classes from \texttt{statsmodels} actually map to a class in an entirely different module \texttt{pandas}.  Furthermore, because of Python packaging, multiple Python classes from within a module appear with different qualified names (as shown by the \texttt{bokeh} classes).

From the seed set of 500 modules that we started with, we ended up with a result set of 1017 modules, 167,872 classes and 164,134 functions.  To cleanse the dataset, we loaded each of 167,872 classes returned by the \texttt{inspect} API in a virtual environment, loaded the class using the name returned by the API, and then noted its actual name when we printed a string representation of the class.  We derived a map of classes to the class they were really aliased to as shown in Table~\ref{classes_map}, which resulted in 92,277 unique classes after aliasing.  We employed a similar approach to alias function names; we loaded 164,134 function names, that ended up aliasing to 91,818 functions. 

\begin{figure*}[htb]
\includegraphics[width=0.78\textwidth]{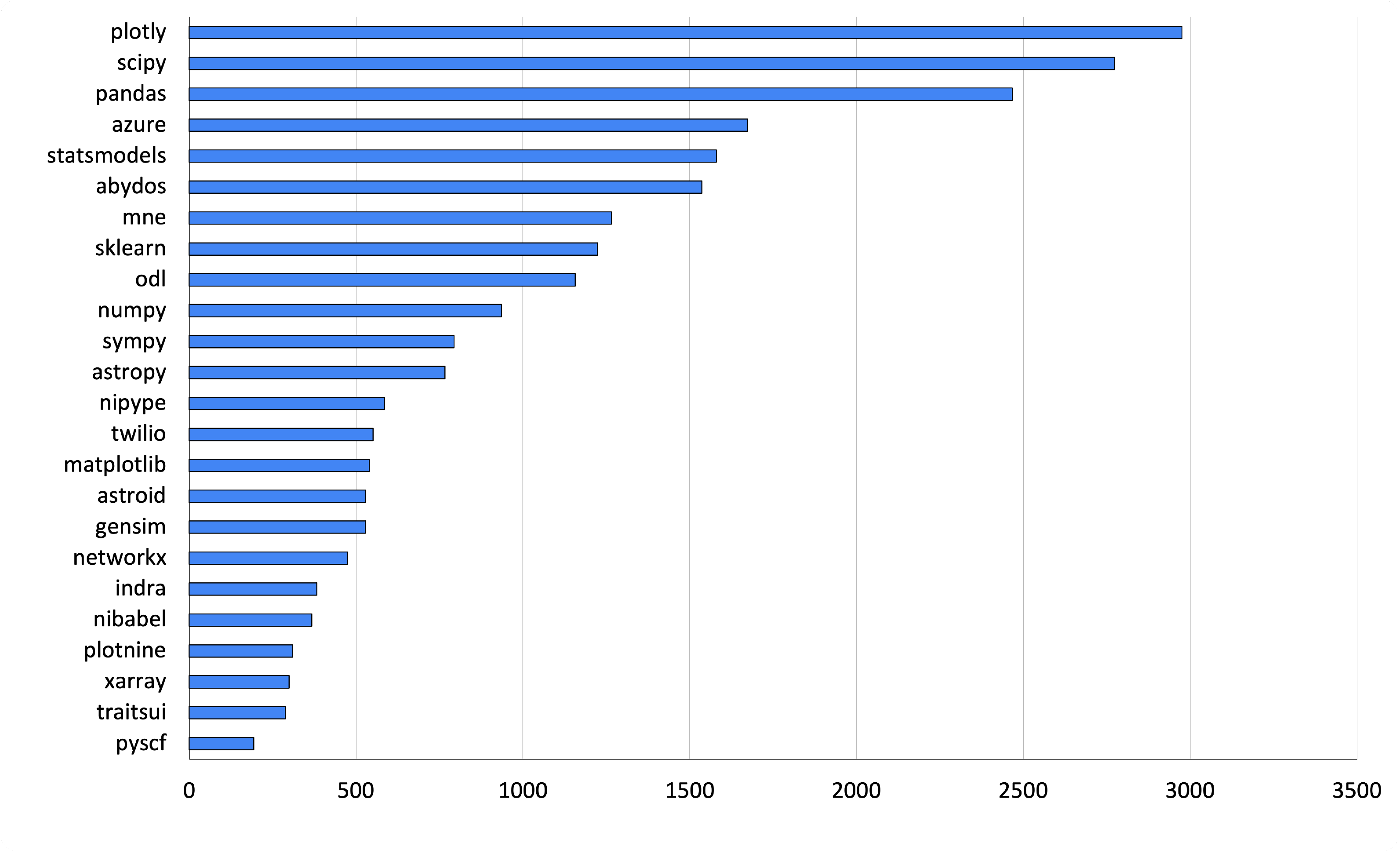}
\caption{Class distribution for top 25 modules}
\label{fig:dataset_summary}
\end{figure*}

Figure~\ref{fig:dataset_summary} describes the distribution of classes in the top 25 modules.  As one can see from the figure, the modules cover a diverse set of functionality; it contains libraries from visualization (e.g. \texttt{plotly}) to cloud management (e.g. \texttt{kubernetes}) to data science libraries (e.g. \textit{sklearn} and \textit{pandas}).
In total we had 26,800 class methods and 53,441 functions with docstrings.  

\section{Type Inference with Docstrings}
\subsection{Extraction of types}
As shown in Figure~\ref{running_example}, documentation in API libraries is well structured, and tends to be written using rich structured text to enable documentation generation from packages like \texttt{Sphinx}.  One question we addressed was how one might leverage information retrieval techniques to infer type information from such documentation.  We focus here on returns, to illustrate our method, as described in Algorithm~\ref{extraction}.  Given a set of modules $l$, we gather all the functions and methods declared in the module into $lf$.  For each $f$, we collect its class (if it is a method) into a set $C$, and get the corresponding docstring $r$.  We use the \texttt{sphinx} library in Python to parse the docstring into restructured text.  In our example, this will strip the `Returns' portion off the entire method docstring, so we have the text shown in Figure~\ref{running_example}.  This structured text $r$ contains each class and function's return value in an informal manner; for instance, in  Figure~\ref{running_example}, it is stated that the return value is either a DataFrame or a TextParser.  To infer the qualified type, we create a `document' $d$ for each function or method, setting the fields of function and content, and index $d$ in an ElasticSearch text index.  At the end of inspection of all modules, every method's return type has been added to the index.  We then loop through all classes in $C$ and search the ElasticSearch index for all $d$ documents that have this class mentioned in them in their return text.  Each $d$ has an inferred type set where the fully qualified classname $c$ is added the function's return type.  Every $d$ in the index at the end of the extraction process is then a function for which we have inferred a type based on docstrings, if $d$ has an inferred type field set.

\subsection{Cleansing}
Because type inference with this mechanism can be quite noisy, we employ a postprocessing step to filter out erroneous annotations. In particular, for a method and its list of inferred types returned from the above step, we perform the following: 
\begin{itemize}
\sloppypar
    \item Using the map of classes to the class they were really aliased to (see Table~\ref{classes_map}), we map each return user-defined type to its correct alias. For example, the class \texttt{pandas.DataFrame} gets mapped to \texttt{pandas.core.frame.DataFrame}. We note that both forms are valid in Python, and in fact, user code will frequently contain imports of \texttt{pandas.DataFrame}, but at runtime the interpreter will return \texttt{pandas.core.frame.DataFrame}.
    \item Remove any type that can not be resolved to any valid type, based on classes that the inspect API provides us, but they fail when one tries to load them at runtime because they do not exist.
    \item Remove user defined types from different libraries, when we have classes as return types which are candidates for the type within the same library.  This last approach is based on the heuristic that if a class with the same name is present in the same library it is more likely to be a candidate for return than a class with the same name from another library.  We note that existing systems for type inference such as TypeWriter~\cite{typewriter} ignore the fully qualified name of the class, which is problematic because we observed this as an issue in our work.
    \item Remove all other classes if a builtin or a primitive is a match.  This step is necessary to avoid matches to classes which have the same name as a builtin or a primitive (e.g., \texttt{Dict}) but clearly are unlikely matches. 
\end{itemize}

\begin{algorithm}
  \caption{Docstring  extraction algorithm}\label{extraction}
  \begin{algorithmic}[1]
    \Procedure{Extract}{$l$}\Comment{The list of modules}
        \State $C$ $\gets$ \texttt{new set}\Comment{set of all classes}
      \For{\texttt{each $m$ in} $l$}
        \State $lf$ $\gets$ \texttt{inspect}($m$)\Comment{list of methods}
        \For{\texttt{each} $f$ in $lf$}
          \State \texttt{add f.class to} $C$ 
            \State $r$ $\gets$ \texttt{inspect}($f$)\Comment{inspect function}
            \State $t$ $\gets$ \texttt{parse}($r$)\Comment{Sphinx to parse text}
            \State $d$ $\gets$ \texttt{new JSON} \Comment{doc for ElasticSearch} 
            \State $d.function \gets f$ 
            \State $d.content \gets t$ 
           \State \texttt{index(d)} \Comment{ElasticSearch index}
       \EndFor
      \EndFor
    \For{\texttt{each $c$ in} $C$}
        \State $ld$ $\gets$ \texttt{search}($c$)\Comment{search for all mentions of c}
       \For{\texttt{each $d$ in} $ld$}
            \State $d.rettype \gets c$ 
       \EndFor
       \EndFor
    \EndProcedure
 \end{algorithmic}
\end{algorithm}

\section{Type Inference with Analysis}
One method to infer types is to perform dataflow over millions of scripts in GitHub, and observe what methods get called on objects returned by a specific method call.  We outline in ~\ref{analysis_challenges} a novel set of changes we introduced into static analysis infrastructure to support this type of analysis.  We then describe how to actually perform duck typing in Section~\ref{duck_typing}.  

\subsection{Extended Analysis Approach}
\label{analysis_challenges}
To perform this dataflow, we confined the scope of our analysis to the level of each Python file in GitHub.  Analysis needs starting points.  We used each method in the script as a starting point, as well as the script itself to ensure maximal coverage of the code in the script.  Our analysis was inter-procedural, so that as shown in Figure~\ref{running_example}a we followed the dataflow into the procedure \texttt{massage\_data} to find that the return value of \texttt{pandas.read\_csv} has both \texttt{dropna} and \texttt{drop} called on it, followed by a call to \texttt{head} guarded by a conditional.

No Python script is self-contained; it always includes imports of libraries and API calls, or user modules with code contained in other files.  
To perform analysis on a large number of files under such circumstances, it was important to not assume that we would be able to create a large number of stubs for such calls, or assume that we could analyze the library code.  We created a mechanism we termed `turtles` to handle such imports or calls on functions that were not part of the script.  The basic approach is that all returns from API calls are represented as instances of a single ``turtle'' type and all calls on such objects return new instances of that type.  Similarly, access to properties of those objects return the object itself.  This can be expressed easily in common analysis frameworks and formalisms, as it requires customization of three aspects of analysis.  We present these three in terms of the analysis abstractions that need to be customized for any analysis framework. We also make an actual implementation available as open source for the larger community\footnote{URL will be provided should the paper be accepted}.

Overall, there are 3 key changes required for any analysis framework to allow a turtle based analysis of the program:
\begin{enumerate}
\item The imports of the required APIs need to be replaced by turtle creations.  The way {\tt import} calls are represented will vary amongst analysis frameworks but in our implementation, we modeled the import call itself directly as a call to a synthetic function that returns a newly-allocated object of ``turtle'' type.  This function is analyzed using call-site sensitivity, i.e. a separate analysis for each call, so that each API import creates a different turtle object.  In Figure~\ref{running_example}, {\tt read\_csv} is imported, so the return of the call on it is represented by a turtle.
\item The semantics of property reads need to be changed so that any property read of a turtle returns the container object itself.  We model this by performing field-insensitive analysis for objects of turtle type, i.e. by modeling all properties of those objects with a single property.  And, when turtle objects are created, we assign the turtle object itself to its single property.
\item The semantics of function calls must be augmented such that any call on an object of turtle type to be a synthetic function that returns a new turtle object.  For function calls, we simply model every function with the same synthetic function that returns a new turtle.  In Python, a call like {\tt pd.read\_csv} consists of first a property read and then a call.  Since property reads on turtles return the same object already, the synthetic model of function calls suffices for method calls too.
\end{enumerate}
\begin{figure}[htb]
\includegraphics[scale=0.4]{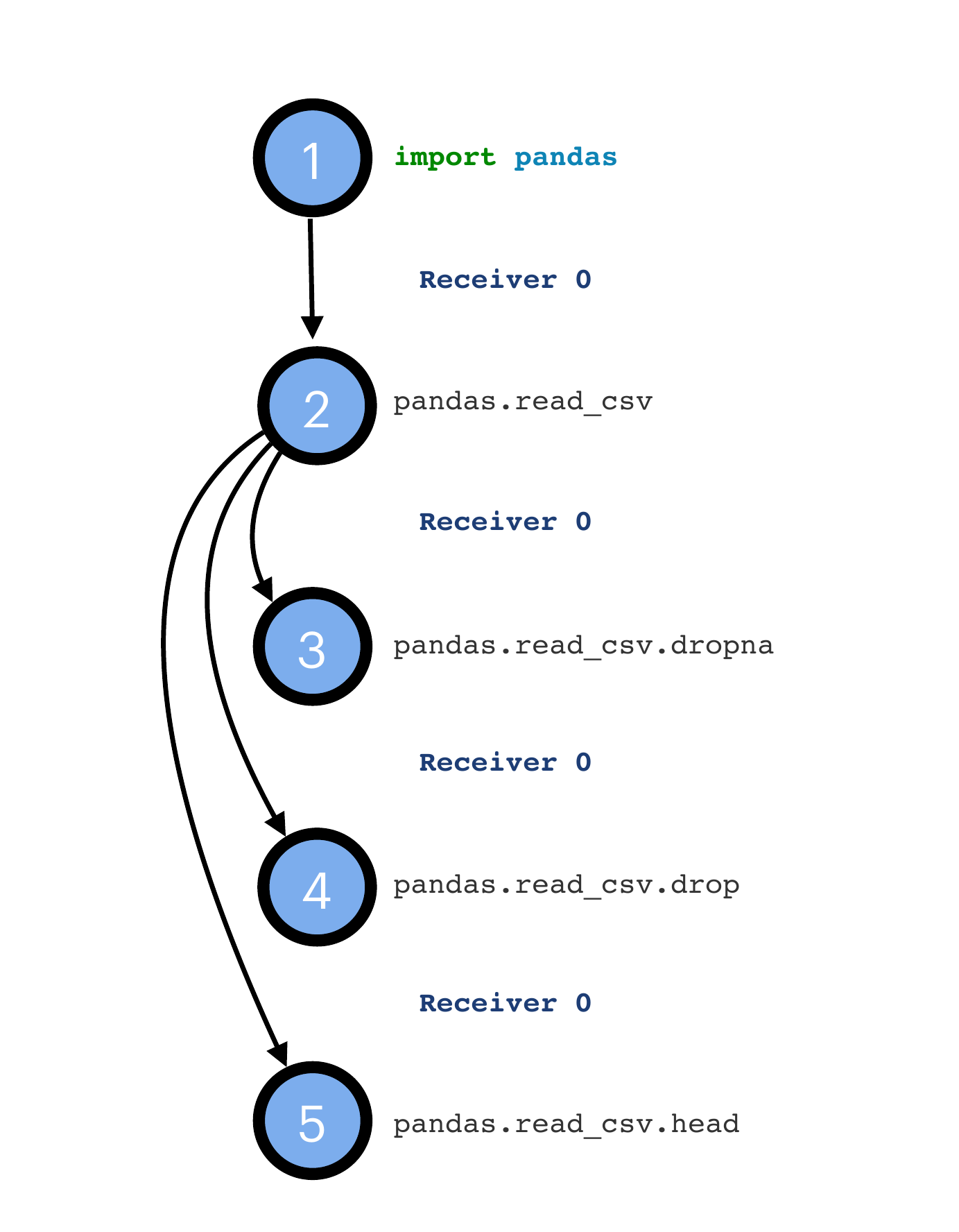}
\caption{Dataflow for script 1}
\label{fig:dataflow}

\end{figure}

\begin{figure*}[htb]
\small
\begin{centering}
\begin{lstlisting}[language=Python,escapechar=|]
... initialization code ...
v40 = invokestatic <Lpandas, import()Lpandas;>   s.py [2:7]->[2:13] [40=[pandas]]
v44 = new <Lscript s.py/massage_data>@21         s.py [2:0]->[14:17] [44=[massage_data]]
global script s.py/massage_data = v44            s.py [2:0]->[14:17] [44=[massage_data]]
putfield v1.< massage_data > = v44               s.py [2:0]->[14:17] [44=[massage_data]]
v47 = getfield < read_csv > v40                  s.py [9:7]->[9:22] [40=[pandas]]
v46 = invokeFunction  v47,v48:#iris.data         s.py [9:7]->[9:35] [46=[data, sample]]
BB2
v51 = invokeFunction  v44,v46 s.py [10:7] -> [10:25] [51=[data, sample]44=[massage_data]46=[data, sample]]
BB3
v56 = invokeFunction  v5,v51 s.py [11:3]->[11:12] [5=[len]51=[data, sample]]
BB4
v55 = binaryop(gt) v56 , v53:#10                  s.py [11:3]->[11:17] [53=[cmp0]]
conditional branch(eq, to iindex=38) v55,v58:#0   s.py [2:0]->[14:17]
BB5
v60 = getfield < head > v51                 s.py [12:13]->[12:22] [51=[data, sample]]
fieldref v61:#n.v58:#0 = v53:#10                  s.py [12:23] -> [12:24] [53=[cmp0]]
v59 = invokeFunction v60 n:53               s.py [12:13]->[12:30] [59=[sample]53=[cmp0]]
BB6
\end{lstlisting}
\caption{IR of script 1 from example code}
\label{code:ann_post}
\end{centering}
\end{figure*}

Our extended analysis framework performs a standard combined call graph construction and pointer analysis that computes, for each call site, what functions may be called, and for each value, what objects it may hold.  Analysis starts at a root function, analyzing each instruction, adding functions found at call sites to a work queue.  To make the workings of the analysis more concrete, we will use the IR for the script of Figure~\ref{running_example}a.  The code is organized as a sequence of basic blocks (denoted {\tt BB0}, {\tt BB1}, etc) of operations such as property reads, and all values are in Static Single Assignment form.  We show how the analysis works for turtles by stepping through what the analysis does when the script is analyzed:
\begin{itemize}
\item instruction 2 is the import corresponding to line 1 of Figure~\ref{running_example}a.  This assigns the imported script to {\tt v40}, which we represent with turtle $t_1$.
\item instructions 3-5 create the inner function {\tt massage\_data} from lines 3 to 8.  Functions are represented as objects in our analysis, since they can be first class.
\item instruction 6 reads the property {\tt read\_csv} from {\tt v40}, which holds the imported {\tt pandas} script, and assigns it to {\tt v47}.  This is also $t_1$.
\item instruction 7 calls {\tt v47} as a function.  Since {\tt v47} holds $t_1$ and the semantics of function calls on turtles is to create a new turtle, we assign the new turtle $t_2$ to {\tt v46}.
\end{itemize} 
The rest of the instructions are mostly analogous, except one
\begin{itemize}
\item instruction 9 calls {\tt v44}, which is {\tt massage\_data}.  This is not a turtle, so the code for that function is added to the work queue of the analysis.  $v46$ is passed as an argument, corresponding to passing the result of the {\tt read\_csv}.

\end{itemize}

There is one aspect of analysis not illustrated by this code snippet: at line 10 of Figure~\ref{running_example}a, the built in {\tt len} call will be passed a turtle returned by {\tt read\_csv} and ultimately {\tt massage\_data}.  Since the analysis makes no assumption about the meaning of a turtle, we treat calls to primitives as simply returning any of the turtles that are used as arguments.

\subsection{Duck Typing}
\label{duck_typing}
As described above, our analysis is neither sound nor complete.  Traditional approaches to duck typing require that for every object $O$ that is returned from a method call $M$, you observe the set of method calls on $O$ which we call $F$, and $F$ must be defined in a given class $C$ in order to infer that $C$ is a return type $M$.  Because we may have imprecision in analysis, it is possible that we have methods in $F$ that are incorrect.  For instance, in Figure~\ref{running_example}, the call to {\tt head} is under an {\tt if}, so it might not be called.  This code would work even if small tables returned a type that did not support \texttt{head}.  To handle this situation, we approximate duck typing by instead computing the size of $F \cup D$ where $D$ is the set of all methods defined for $C$.  The likelihood that type inference was correct is governed by two factors: (a) the size of $F \cup D$, and (b) the number of classes that are possible types for a given method return value.  Clearly, as (a) increases, confidence in type inference grows.  However, a small number of classes in (b) in combination with a small number of shared methods in (a) can sometimes still imply a valid inference.  An example of such a case is shown in Table~\ref{classes_example}, where, for instance, we find that \texttt{pandas.array} returns \texttt{pandas.core.arrays.base.ExtensionArray} correctly, and in fact \texttt{pandas.core.arrays.sparse.array.SparseArray} is a subclass of \texttt{pandas.core.arrays.base.ExtensionArray}.  We discuss how to cleanse the types next.

\begin{table}[htb]
    \caption{Example of sharing small number of methods}
    \label{classes_example}
    \begin{tabular}{l|l|l}
    \toprule
      Method & Return & \# \\
      \midrule
pandas.array & pandas.core.arrays.base.ExtensionArray & 2 \\
pandas.array & pandas.core.base.IndexOpsMixin & 2 \\
pandas.array & pandas.core.arrays.sparse.array.SparseArray & 1 \\
     \bottomrule
    \end{tabular}
\end{table}

\subsection{Analysis Cleansing}
\label{analysis_cleansing}

We often find a large number of spurious types from our initial duck typing of code, and we filter them in a series of steps:
\begin{itemize}
    \item Since our duck typing is not entirely precise, the first step is to filter candidates types to those that match the largest number of methods called in the code.
    \item There are often many concrete types that share a common supertype that is also present in the set of types.  In this case, we remove the subtypes, since they are covered by the supertype.
    \item Sometimes most of the types in a set share a supertype $S$ that is not itself in the set.  In this case, we remove types that are not subtypes of $S$, since they are often due to analysis imprecision.
    \item We use lists of functions and classes to remove items that are in fact modules, but appear ambiguous due to the fact imports can be of anything.
    \item We eliminate classes and functions that were not valid as before, and use their aliases.
\end{itemize}

\section{Evaluation}




\subsection{How precise are labeled types?}
\subsubsection{Evaluation against dynamic types}
\begin{table}[]
    \caption{Summary of number of passed and failed tests and number of methods inferred for each module}
    \label{dynamic_type_packages}
    \begin{center}
    \begin{tabular}{l r r r}
    \toprule
         Module & Passed & Failed & Methods  \\
         \midrule
         Flask & 408 & 29 & 255 \\
         Numpy & 4,881 & 5,139 & 262 \\
         Scikit-learn & 17,900 & 1,142 & 371 \\
         Sympy & 1,553 & 214 & 785 \\
         Pandas & 48,625  & 6,453 & 624 \\
         \bottomrule
    \end{tabular}
    \end{center}
\end{table}

To develop a gold standard for our evaluation, we collected a set of types by observing their runtime types.  We targeted 5 repositories from our set of 408 repositories that (a) used \texttt{pytest} for unit testing, (b) seemed to be set up relatively easily without a set of additional dependencies on databases, servers etc.  For each function invoked by \texttt{pytest} in the tests, we inserted a wrapper function which would log its return type before return.  We leveraged monkey patching in \texttt{pytest} and \texttt{pytest} fixtures to insert our wrapper.  Table~\ref{dynamic_type_packages} shows the number of tests that passed or failed in each package\footnote{We note that monkey patching caused a larger number of test failures, for various reasons which were not easy to fix.}.  The types gathered by monkey-patching are always sound, but not necessarily complete.  We gathered the 2284 distinct methods for which we had types.  

Each method was annotated often with multiple types. 
We manually inspected some of the cases and augmented the set of dynamic types when we could based on documentation, and running the code. We tried to extend this beyond the libraries in Table~\ref{dynamic_type_packages} but for many libraries we tried, it was not possible to set tests up even when they did support \texttt{pytest}. 

\begin{table}
\caption{Statistics about dynamic types found}
\label{basic_stats}
\begin{tabular}{lll}
\toprule
                             & \parbox{.7in}{\raggedright Number of dynamic types found} & \parbox{.7in}{\raggedright Mean number of types}  \\
\midrule
PyType                       & 105    & 1.0    \\
Docstrings                   & 168     &   1.125 \\
Static Analysis              & 132     & 1.96 \\
Docstring + analysis        & 203  & 1.69 \\
\bottomrule
\end{tabular}
\end{table}

Table~\ref{basic_stats} shows the total number of tests we ran to get the 2,284 methods we gathered from dynamic typing.  The number of matches were quite low for each type inference technique, but investigation showed that this was because we frequently received from the runtime, method names for functions that dropped the class name\footnote{We used \texttt{dill} to get the fully qualified name of a function in logging return types.}. Of the three methods, extraction from docstrings retrieved the types for most functions. Analysis was next followed by \texttt{PyType}.  Docstrings when combined with analysis yielded return types for 203 methods, which is ~9\% of the methods we had dynamic information for.

\begin{table}
\caption{Precision of docstrings and analysis based type inference versus PyType}
\label{precision_summary}
\centering
\begin{tabular}{llll}
\toprule
                             & Precision & Recall & F1-score  \\
\midrule
PyType                       & 0.067     & 0.067  & 0.067     \\
Docstrings                   & 0.661     & 0.529   & 0.587     \\
Static Analysis              & 0.489     & 0.549  & 0.517     \\
\bottomrule
\end{tabular}

\end{table}

Table~\ref{precision_summary} shows the results of precision and recall for \texttt{PyType}, and separately for type inference based on docstrings and type inference based on static analysis and duck typing.  \texttt{PyType}'s $F1$ score was very surprisingly low ($.067$), but this result is in consistent with the 6.1\% accuracy reported by \cite{10.1145/3426422.3426981} when the type is a user defined type.  
In contrast, the $F1$ score for type inference based on docstrings was $0.587$, and $ 0.517$ with static analysis; a significant improvement over  \texttt{PyType}.  

\subsubsection{Evaluation of class constructors}
Dynamic typing is one method to analyze the precision of our type inference.  We exploit a feature of the Python language as a type of sanity test for the precision of static analysis based type inference.  In Python, as in many dynamic languages, a constructor is simply another method.  We used this fact to generate a gold standard of methods for which we know the return type.  We gathered all classes 92,277 classes from \texttt{inspect}, and asked about whether their constructors were inferred correctly by our technique for type inference using static analysis\footnote{This technique cannot be applied to test the precision of the method which infers types based on docstrings because init methods do not have docstrings which specify return values correctly.}.  Recall for constructors was $0.0459$, indicating that only a small percentage of classes were used in practice.  Of those, static analysis based duck typing produced the correct type for 4,236 types, and an incorrect value for 130 types, for a precision that was $0.97$.  The errors were due to errors in gathering class definitions.  As an example \texttt{QtNetwork.QLocalSocket} is a class that we see in usage, and it has a method \texttt{waitForConnected} called on it in code.  However, in the \texttt{inspect} output, no method \texttt{waitForConnected} was found, and hence it was not associated with any class.  We note that in general, the \texttt{inspect} API from Python had several inaccuracies which added noise to the process.  Nevertheless, the test with class constructors suggests the analysis and duck typing approach does work.

\subsubsection{Manual annotation}
To evaluate the type inference for the two techniques further, we selected a random sample of methods for each technique, and we tried to manually evaluate if the return type was correct.  Note in this case, we cannot actually evaluate recall or F1, but this sort of qualitative assessment is useful to understand where the weaknesses of each method are.  For analysis, we tried to find as much information as we could from documentation on the web or what we had gleaned from inspection to make the decision on whether the returned type was correct or not.  

\sloppypar
\noindent  \textbf{Static Analysis Sample:} For 25/108 methods, we could not find enough documentation to infer the return type correctly.  For the remaining methods, we often returned multiple types.  
Across all those returned types, we were correct on 71/163 (43.56\%) cases (where each case reflects a specific type inference), which is lower than what we observed with dynamic typing, which may just reflect sampling noise.  One observation from this exercise is that we often find classes that are conceptually very similar but they are not related from a type perspective.  As an example, we found \texttt{scipy.spatial.kdtree.KDTree} as a return type for \texttt{sklearn.neighbors.BallTree}.  Both are conceptually related, both are derived from \texttt{BinaryTree}, but of course one cannot be substituted for another.  This is a weakness of the duck typing approach in general.  

\noindent \textbf{Docstrings Sample:} We created another random sample of 200 methods from docstrings type annotations. We could not manually verify the return type of 67 methods which were mostly internal setter functions inside libraries like \textit{plotly}. For the rest of the methods, we predicted the return type correctly for 103/133 (77\%). One common issue with docstring-based types is its impreciseness when the documentation is not sufficient or vague. In \texttt{numpy} for instance, documentation would frequently state that the return value is an \texttt{array}, but what was being returned was \texttt{numpy.ndarray}.  In such cases, relying on usage patterns could infer better types. 


\subsection{Weaknesses of static typing in \texttt{PyType}}
The next question we evaluated was whether our methods for type inference addressed some of the weaknesses we referred to in the Introduction with static typing tools such as \texttt{PyType}.  Similar to \cite{10.1145/3426422.3426981}, we chose to compare against \texttt{PyType} because of the observation that \texttt{PyType} is slightly better than \texttt{MyPy} in type inference.

\begin{figure}
\centering
\includegraphics[width=0.5\textwidth]{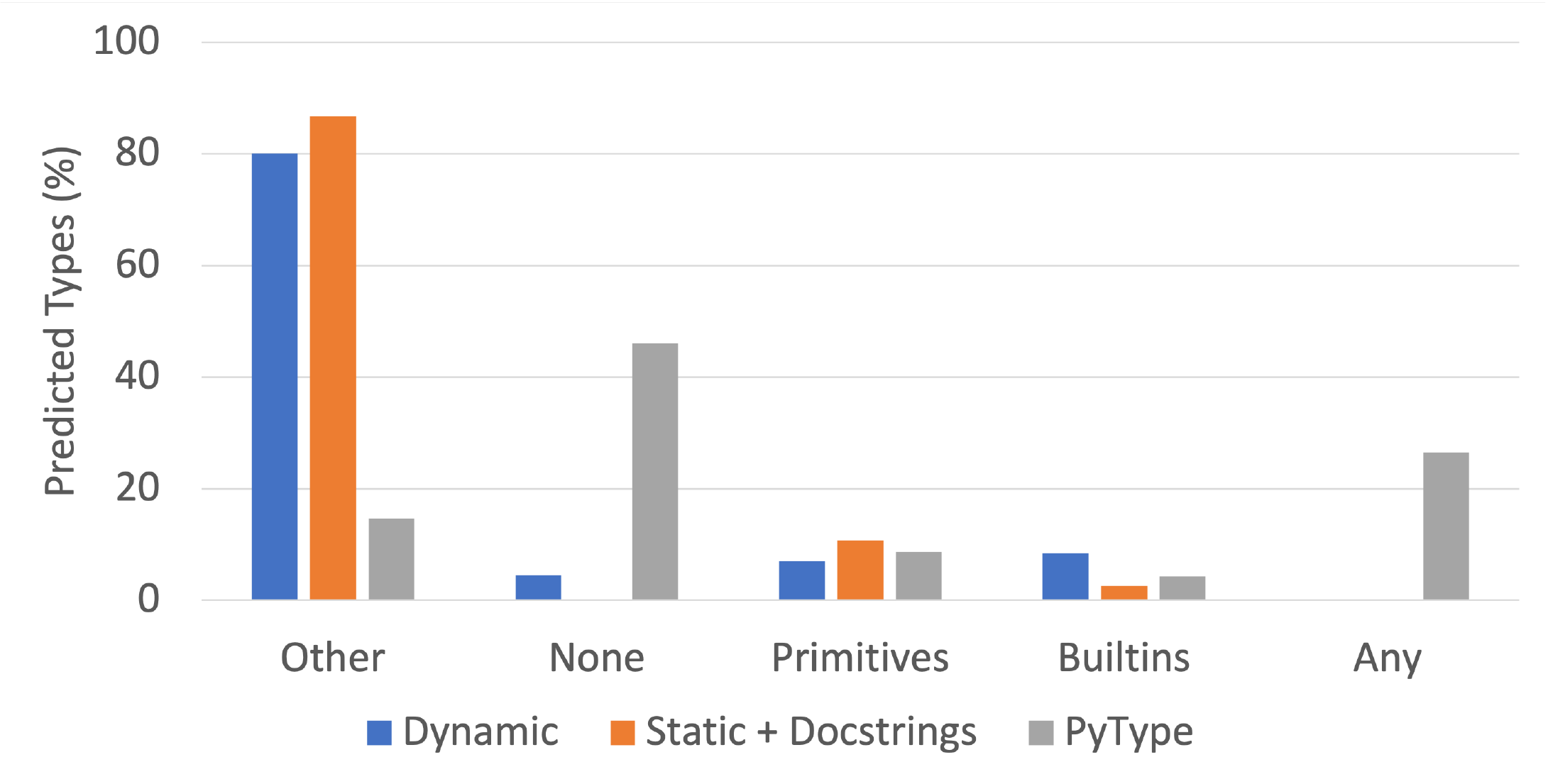}
\caption{Percentage of predicted types using Dynamic Typing, PyType and our approach}
\label{fig:pytype_dynamic_us}
\end{figure}

Figure~\ref{fig:pytype_dynamic_us} shows the distribution of types for dynamic typing, versus \texttt{PyType} and our methods on this gold standard we developed.  Once again, as we discussed in Figure~\ref{fig:pytype}, \texttt{PyType} tends to produce less user defined types, and produces a large percentage of types that are labeled \texttt{Any} which is not very precise type information.  Our method is biased against void types, unless we infer those from documentation.  For the purposes of harvesting high quality labeled data, it is less important to model `void' correctly.  For all other categories, we seem to infer as many types as produced by dynamic types. 

\begin{table}[]
    \caption{Confusion matrix for PyType Against Dynamic types}
    \label{pytype_confusion_matrix}
    \begin{tabular}{lllllll}
    \toprule
     Dynamic   & Primitive & None & Any & BuiltIn & Class \\
     \midrule
     Class     & 0 & 0 & 270  & 0 & 0 \\
     Builtin   & 0 & 0 & 9    & 4 & 3 \\
     Primitive & 3 & 0 & 17   & 0 & 0 \\
     None      & 0 & 4 & 1    & 0 & 0 \\
     \bottomrule
    \end{tabular}
\end{table}

\begin{table}[]
    \caption{Confusion matrix for Static Analysis- the number for Class reflects errors (correct answers is in parentheses)}
    \label{analysis_confusion_matrix}
    \begin{tabular}{llll}
    \toprule
     Dynamic   & Primitive & BuiltIn & Class \\
     \midrule
     Class      & 0 & 0 & 124  (\textbf{141}) \\
     Builtin    & 0 & 0  &  13 \\
     Primitive  & 2 & 0 & 3 \\
  
      \bottomrule
    \end{tabular}
\end{table}

\begin{table}[]
    \caption{Confusion matrix for Docstrings - the number for Class reflects errors (correct answers is in parentheses)}
    \label{docstrings_confusion_matrix}
    \begin{tabular}{llll}
    \toprule
     Dynamic   & Primitive  & BuiltIn & Class\\
     \midrule
     Class   & 24 & 1 &  56 (\textbf{134}) \\
     Builtin & 0 & 1 & 2  \\
     Primitive & 23 & 0 & 0 \\
     \bottomrule
    \end{tabular}
\end{table}

To examine the nature of each typing method, and its errors against the dynamic types, we computed a confusion matrix for each method.  Table~\ref{pytype_confusion_matrix} shows the same behavior as observed in Figure~\ref{fig:pytype} for \texttt{PyType}, and we note that tendency to respond with \texttt{Any} in this system is true across all types, but exacerbated for user defined types, to the point where none of the user defined classes were ever inferred correctly. In fact, \texttt{PyType} frequently returned the name of a module (e.g., \texttt{sympy} for user defined classes such as \texttt{sympy.core.power.Pow}).  The inference techniques we used had the exact opposite bias.  Table~\ref{analysis_confusion_matrix} shows that analysis tends to err on the side of providing user defined types.  The confusion for builtins reflects coarseness in how we modeled flow - if some object was retrieved from a tuple or a list and then a method was called on the object, we falsely assumed a direct data flow.  Table~\ref{docstrings_confusion_matrix} shows the confusion matrix for docstrings, which also shows a similar error pattern as analysis, frequently confusing primitives and built-ins with user defined types.  Most of those errors in the docstring case came from the fact that the docstring for \texttt{numpy} methods frequently return a user defined class \texttt{numpy.bool\_} but state that they return a \texttt{bool} type.  Similarly for builtin, when tuples were returned but the documentation stated the types being returned in the tuple, we incorrectly stated that the return type was one of the mentioned types.  

We also examined to what extent our techniques and \texttt{PyType} agree on the types returned from static analysis, as shown in Table~\ref{pytype_analysis_confusion_matrix}.  The agreement is small (22\%) even when \texttt{PyType} returns a type that looks like a class, as shown in Table~\ref{pytype_analysis_confusion_matrix}.  As we discuss shortly this agreement is much worse than agreement between our two techniques (61\%).  In many cases, \texttt{PyType} does not return a fully qualified name of the class, so our measure of the overlap of 47 cases was adjusted to consider cases when the class name matched. In 36 cases, \texttt{PyType} returned a module as a returned type, which means in 36/209 cases (17\%) \texttt{PyType} is returning imprecise information about types when it infers a class.  

\begin{table}[]
    \caption{Confusion matrix for PyType vs. Static Analysis- the number for Class reflects differences (agreement in parentheses)}
    \label{pytype_analysis_confusion_matrix}
    \begin{tabular}{llll}
    \toprule
     Pytype   & Primitive & BuiltIn & Class \\
     \midrule
     Any        & 0 &  0 & 658 \\
     Class      & 0  & 0 & 162  (\textbf{47}) \\
     Builtin    & 0 &  0 & 100 \\
     Primitive  & 2 & 0 & 1 \\
      \bottomrule
    \end{tabular}
\end{table}

A similar comparison with type inference based on docstrings is shown in Table~\ref{pytype_docstrings_confusion_matrix}.  When \texttt{PyType} produced a class, it matched a docstring based class in 47 cases; with an agreement of about 38\%.  Docstring based inference seemed especially prone to disagreeing with \texttt{Pytype} on builtins, most likely because documentation often refers to both the data structure and the types held in it (e.g. \texttt{list of int}.  This is currently a weakness of docstring extraction that could be addressed with better natural language processing techniques, but this is for future work.  Once again, \texttt{PyType} returned a module instead of a class 34 times, which is 34/124 (27\%).

\begin{table}
    \caption{Confusion matrix for PyType vs Docstrings - the number for Class reflects differences (agreement in parentheses)}
    \label{pytype_docstrings_confusion_matrix}
    \begin{tabular}{llll}
    \toprule
     PyType   & Primitive & BuiltIn & Class\\
     \midrule
     Any     & 617 & 183 & 1051 \\
     Class   & 36  & 17 &  77 (\textbf{47}) \\
     Builtin & 133 & 124 & 66  \\
     Primitive & 247 & 0 & 14 \\
     \bottomrule
    \end{tabular}
\end{table}

\begin{figure}[htb]
    \includegraphics[width=0.45\textwidth]{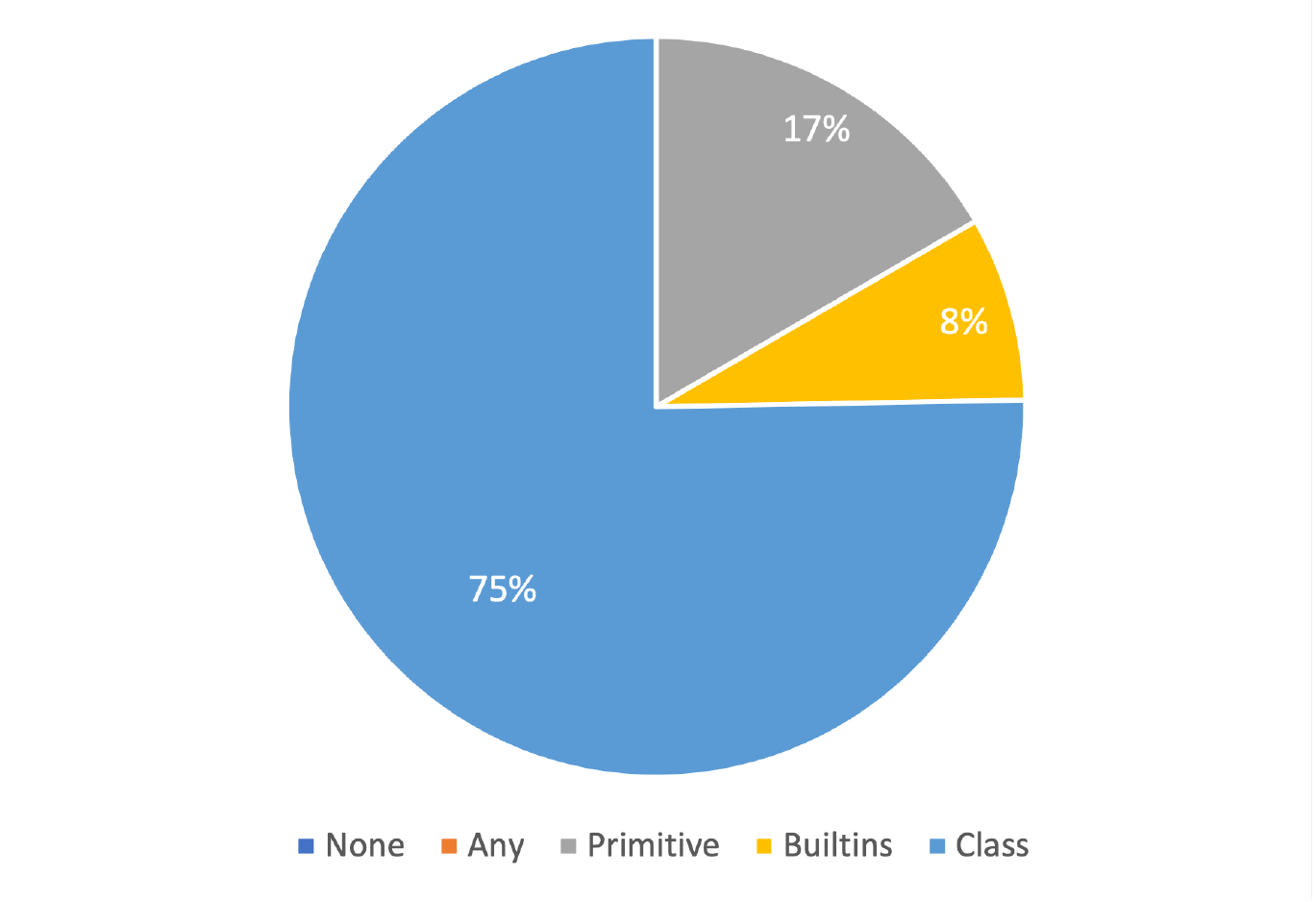}
\caption{Static Analysis and Docstrings-based predictions per type category}
\label{fig:our_categories}
\end{figure}

\subsection{Properties of the inferred types dataset}
Table~\ref{summary_both_types} shows some summary statistics of the two methods of type inference.  As shown in the table, together the two techniques yield over 37,000 labeled types.  The degree of intersection between the two was small (410) because the focus of each is quite different.  When they did produce types for the same methods, they agreed in 249/410 cases (61\%).  We also show in Figure~\ref{fig:our_categories} the distribution of predictions per category type. Compared to PyType (see Figure~\ref{fig:pytype}, our approach clearly complements \texttt{PyType} by producing more user defined types instead of \textit{None} and \textit{Any} types from \texttt{PyType}. This and the accuracy results shown earlier support the claim that our work can indeed produce better quality type annotations (turning inconclusive types like \textit{Any} and \textit{None} to real types) that can further improve existing type inference techniques.   

\begin{table}
\caption{Summary of types inferred by the two methods}
\label{summary_both_types}
\begin{tabular}{lrr}
\toprule
                             & Total & Average types per method \\
\midrule
Docstrings                   & 22,041    &  1.12   \\
Static Analysis              & 15,486     & 1.50     \\
\bottomrule
\end{tabular}
\end{table}

\begin{figure*}
\includegraphics[width=0.78\textwidth]{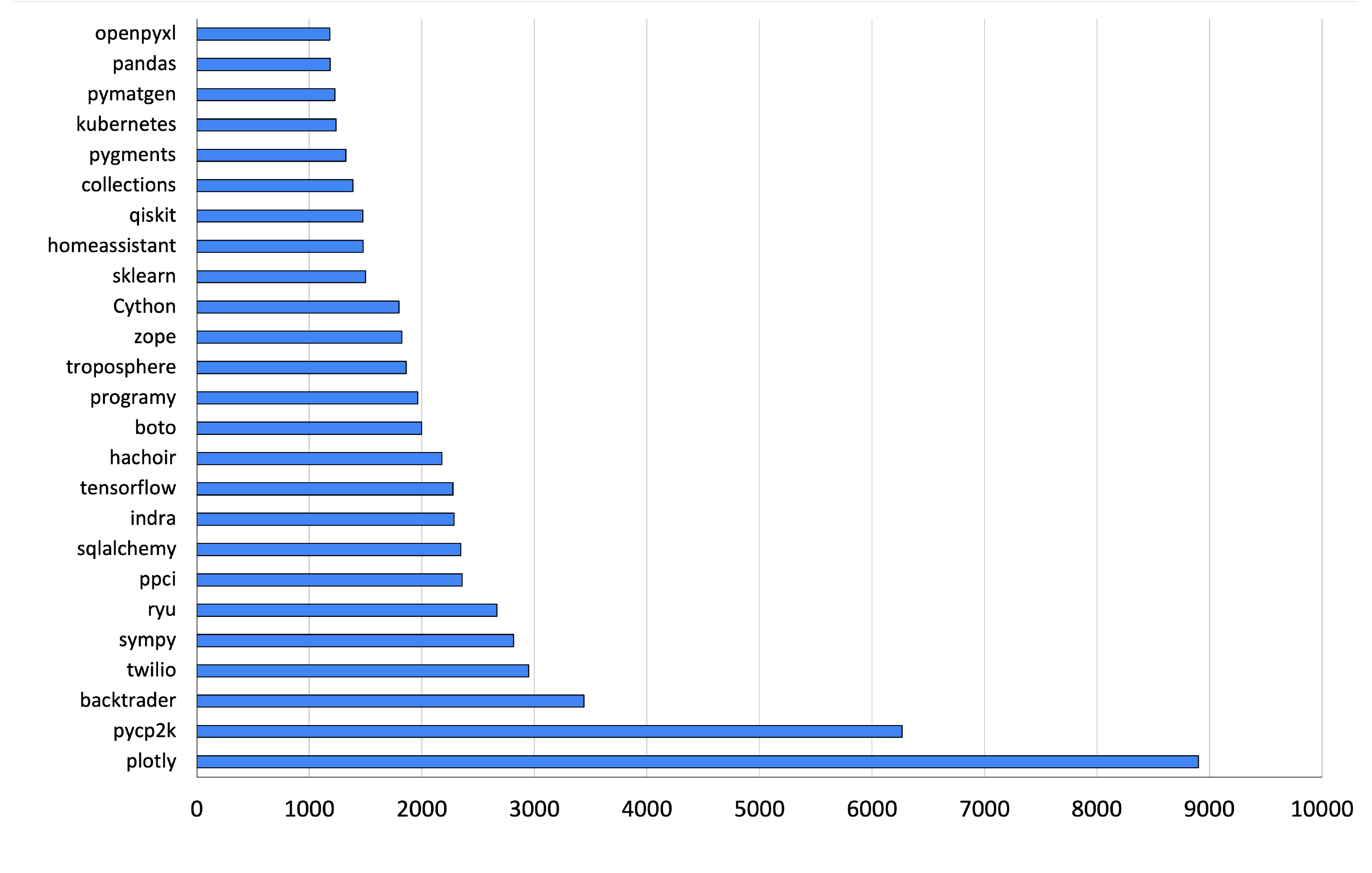}
\caption{Distribution of inference for top 25 modules}
\label{fig:disbn25}
\end{figure*}

Figure~\ref{fig:disbn25} shows the distribution of the top 25 modules for which the two methods inferred types. Some of these modules had the most classes as shown in Figure~\ref{fig:dataset_summary}, but not all.  This is in keeping with the fact that the two techniques have different strengths, so modules with a larger number of classes do not completely govern the effectiveness of type inference.

\begin{table}
\centering
\small
\caption{Top-20 predicted types compared to TypeWriter}
\label{typewriter_us}
\begin{tabular}{rlll}
\toprule
\multicolumn{1}{l}{} & TypeWriter     & This Work &                                                       \\
\midrule
1                    & None           & list~    &                                                       \\
2                    & Unknown        & float~   &                                                       \\
3                    & str            & str~     &                                                       \\
4                    & bool           & array~   &                                                       \\
5                    & int            & bool~    &                                                       \\
6                    & Any            & \multicolumn{2}{l}{plotly.basedatatypes.BasePlotlyType~}         \\
7                    & Response       & int~     &                                                       \\
8                    & dict           & \multicolumn{2}{l}{pandas.core.frame.DataFrame~}                 \\
9                    & Tensor         & \multicolumn{2}{l}{pandas.core.series.Series~}                   \\
10                   & Optional[str]  & \multicolumn{2}{l}{tensorflow.python.ops.variables.Variable~}    \\
11                   & Dict[str, Any] & \multicolumn{2}{l}{tensorflow.python.ops.variables.VariableV1~}  \\
12                   & List[str]      & \multicolumn{2}{l}{scipy.sparse.base.spmatrix~}                  \\
13                   & float          & \multicolumn{2}{l}{numpy.ndarray~}                               \\
14                   & ndarray        & \multicolumn{2}{l}{mne.io.fiff.raw.Raw~}                         \\
15                   & bytes          & \multicolumn{2}{l}{pandas.core.indexes.base.Index~}              \\
16                   & CommonTestData & \multicolumn{2}{l}{mne.epochs.Epochs~}                           \\
17                   & HttpResponse   & \multicolumn{2}{l}{scipy.sparse.data.\_data\_matrix~}            \\
18                   & Dict           & \multicolumn{2}{l}{numpy.ma.core.MaskedArray~}                   \\
19                   & DataFrame      & \multicolumn{2}{l}{sympy.core.expr.Expr~}                        \\
20                   & list           & \multicolumn{2}{l}{mne.evoked.Evoked~}                          \\
\bottomrule
\end{tabular}
\end{table}

\subsubsection{Comparison with TypeWriter}
We also considered evaluating our type predictions against TypeWriter~\cite{typewriter}; a recent neural model that leverages code and the natural language tokens around it. However, we found it hard to run TypeWriter on our datasets since the pre-trained model was not given and TypeWriter does not output the method qualified name, rather it simply names a source file and line number. Therefore, we only focused on an overall analysis of the output of TypeWriter against our predicted type categories. On its test set, TypeWriter outputs the following categories of \texttt{Any} and \texttt{Unknown} (12\%), \texttt{None} (38\%),  primitives (18\%), built-ins (2\%) and 30\% for the rest.  Note that the production of \texttt{Any} reflects again the reliance of such systems on labeled data from static typing tools such as \texttt{MyPy}.  With the caveat that our results are on a different dataset, our results show in Figure~\ref{fig:our_categories} shows that our work produces more user-defined types (75\%).  Table~\ref{typewriter_us} shows the top types reported by TypeWriter compared to our technique.  As one can see. we produce more user defined types in the top 20 set, and our types are fully qualified which is helpful for inference.


\section{Related Work}


In many languages, type inference is a well understood problem.  For statically-typed languages, it is a convenience to reduce the amount of typing at the keyboard to get typing of the program.  Recent versions of Java, for instance, do significant amounts of type inference in the context of generics and lambda functions \cite{0.5555/26369917}.  Historically, languages such as ML \cite{10.5555/549659,Sulzmann00} pioneered type inference such that writing types was usually unnecessary, except for some tricky polymorphic cases.

 Type inference for dynamic languages has always been more approximate.  In part, that is because the dynamic nature of the language permits flexibility such as the same variable having different types in different places; in fact, this is one place where \texttt{MyPy} can have trouble.  Type analysis has been done in Python, but most of the approaches developed so far cannot produce high quality labeled data on a large scale.  Some approaches handle restricted versions of the language (e.g. \cite{maia_moreira_reis_2011}), but these approaches have limited applicability for producing large scale unsupervised data from our large collection of public Python code; we have no control over that code, and it uses all aspects of the language freely.
 
 There have been some approaches for traditional inference, but they all suffer from the extreme difficulty of doing precise analysis in Python, and blur the distinction between type inference and program analysis.  Fritz and Hage \cite{fritz_hage_2017}
 implement type inference via abstract interpretation and
 experiment with tuning the precision by modifying flow sensitivity and context sensitivity.  Moving fully into more traditional static program analysis, Dolby et al. \cite{dolby_et_al_2018} use type inference based on the WALA program analysis framework to find bugs in
 Python-based deep learning code by inferring tensor shapes
 and dimensions.  Beyond purely static analysis, Hassan et al.  \cite{hassan_et_al_2018} implement
 type inference via a MaxSMT solver, maximizing optional
 equality constraints while satisfying all mandatory type constraints.
 
 More approximate techniques have been tried as well.  For instance, Xu et al. \cite{xu_et_al_2016} augment standard type rules with a probabilistic approach to make use of additional information such as identifier naming conventions. While not sound like more traditional type inference, this can yield additional results.  
 
 For all these techniques, they are not designed for our use case.  They will not scale to the enormous collection of libraries that our collection of code uses, and so we attempt to collect information at the boundary of these libraries with duck typing on the application side and gathering information from documentation for the API side. 
 
 Beyond that, machine learning approaches are gaining popularity.  TypeWriter~\cite{typewriter} trained a neural model using a corpus of code, with labeled data derived from user annotations. The trained model is used to predict likely types which are used by feedback directed search and a static type checker (\texttt{MyPy}) to find consistent type assignments.  TypeWriter does consider comments in code, as we do, but as inputs to the neural model.  A key problem with TypeWriter's type inference is that the types the system infers are not qualified.  For user specified types, this is a serious problem because many names are re-used across libraries (e.g., \texttt{numpy.ndarray} is used heavily by several libraries, so returning \texttt{ndarray} alone is unhelpful).  Another neural system Typilus~\cite{typilus} represents code as a graph and trains a Graph Neural Network to find type embeddings. The labeled data it uses are from \texttt{PyType} and \texttt{MyPy}, which have serious shortcomings, because they often produce the type \texttt{Any} for most user defined types.  

Other dynamic languages face a similar issue, and machine learning has been employed for them too.  DeepType~\cite{deeptype} and NL2Type~\cite{nl2type} build deep learning models for JavaScript type annotations.  However, as shown in ~\cite{10.1145/3426422.3426981}, these are sparsely populated in Python repositories, and hence it would be hard to apply this line of work directly to Python.  Our work aims for producing high quality annotated data in an unsupervised manner to allow for building better deep learning models.



\section{Conclusions and Future work}

In this work, we have shown that one can leverage documentation as well as usage information to produce reasonably high quality labeled data for Python at a large scale.  Our techniques achieve significantly better performance than static type checkers. It also produces high quality labeled data, enabling better probabilistic type inference systems.  A next step for future work is to leverage these large scale type annotated data for building better neural models for type inference.

\bibliographystyle{ACM-Reference-Format}
\bibliography{acmart}


\begin{thebibliography}{15}


\ifx \showCODEN    \undefined \def \showCODEN     #1{\unskip}     \fi
\ifx \showDOI      \undefined \def \showDOI       #1{#1}\fi
\ifx \showISBNx    \undefined \def \showISBNx     #1{\unskip}     \fi
\ifx \showISBNxiii \undefined \def \showISBNxiii  #1{\unskip}     \fi
\ifx \showISSN     \undefined \def \showISSN      #1{\unskip}     \fi
\ifx \showLCCN     \undefined \def \showLCCN      #1{\unskip}     \fi
\ifx \shownote     \undefined \def \shownote      #1{#1}          \fi
\ifx \showarticletitle \undefined \def \showarticletitle #1{#1}   \fi
\ifx \showURL      \undefined \def \showURL       {\relax}        \fi
\providecommand\bibfield[2]{#2}
\providecommand\bibinfo[2]{#2}
\providecommand\natexlab[1]{#1}
\providecommand\showeprint[2][]{arXiv:#2}

\bibitem[\protect\citeauthoryear{Abdelaziz, Dolby, McCusker, and
  Srinivas}{Abdelaziz et~al\mbox{.}}{2021}]%
        {abdelaziz2021toolkit}
\bibfield{author}{\bibinfo{person}{Ibrahim Abdelaziz}, \bibinfo{person}{Julian
  Dolby}, \bibinfo{person}{Jamie McCusker}, {and} \bibinfo{person}{Kavitha
  Srinivas}.} \bibinfo{year}{2021}\natexlab{}.
\newblock \showarticletitle{A toolkit for generating code knowledge graphs}. In
  \bibinfo{booktitle}{\emph{Proceedings of the 11th on Knowledge Capture
  Conference}}. \bibinfo{pages}{137--144}.
\newblock


\bibitem[\protect\citeauthoryear{Allamanis, Barr, Ducousso, and Gao}{Allamanis
  et~al\mbox{.}}{2020}]%
        {typilus}
\bibfield{author}{\bibinfo{person}{Miltiadis Allamanis},
  \bibinfo{person}{Earl~T Barr}, \bibinfo{person}{Soline Ducousso}, {and}
  \bibinfo{person}{Zheng Gao}.} \bibinfo{year}{2020}\natexlab{}.
\newblock \showarticletitle{Typilus: neural type hints}. In
  \bibinfo{booktitle}{\emph{Proceedings of the 41st acm sigplan conference on
  programming language design and implementation}}. \bibinfo{pages}{91--105}.
\newblock


\bibitem[\protect\citeauthoryear{Dolby, Shinnar, Allain, and Reinen}{Dolby
  et~al\mbox{.}}{2018}]%
        {dolby_et_al_2018}
\bibfield{author}{\bibinfo{person}{Julian Dolby}, \bibinfo{person}{Avraham
  Shinnar}, \bibinfo{person}{Allison Allain}, {and} \bibinfo{person}{Jenna
  Reinen}.} \bibinfo{year}{2018}\natexlab{}.
\newblock \showarticletitle{Ariadne: Analysis for Machine Learning Programs}.
  In \bibinfo{booktitle}{\emph{Workshop on Machine Learning and Programming
  Languages (MAPL)}}. \bibinfo{pages}{1--10}.
\newblock
\urldef\tempurl%
\url{http://doi.acm.org/10.1145/3211346.3211349}
\showURL{%
\tempurl}


\bibitem[\protect\citeauthoryear{Fritz and Hage}{Fritz and Hage}{2017}]%
        {fritz_hage_2017}
\bibfield{author}{\bibinfo{person}{Levin Fritz} {and} \bibinfo{person}{Jurriaan
  Hage}.} \bibinfo{year}{2017}\natexlab{}.
\newblock \showarticletitle{Cost versus Precision for Approximate Typing for
  Python}. In \bibinfo{booktitle}{\emph{Workshop on Partial Evaluation and
  Program Manipulation (PEPM)}}. \bibinfo{pages}{89--98}.
\newblock
\urldef\tempurl%
\url{https://doi.org/10.1145/3018882.3018888}
\showURL{%
\tempurl}


\bibitem[\protect\citeauthoryear{Gosling, Joy, Steele, Bracha, and
  Buckley}{Gosling et~al\mbox{.}}{2014}]%
        {0.5555/26369917}
\bibfield{author}{\bibinfo{person}{James Gosling}, \bibinfo{person}{Bill Joy},
  \bibinfo{person}{Guy~L. Steele}, \bibinfo{person}{Gilad Bracha}, {and}
  \bibinfo{person}{Alex Buckley}.} \bibinfo{year}{2014}\natexlab{}.
\newblock \bibinfo{booktitle}{\emph{The Java Language Specification, Java SE 8
  Edition} (\bibinfo{edition}{1st} ed.)}.
\newblock \bibinfo{publisher}{Addison-Wesley Professional}.
\newblock
\showISBNx{013390069X}


\bibitem[\protect\citeauthoryear{Hassan, Urban, Eilers, and M{\"u}ller}{Hassan
  et~al\mbox{.}}{2018}]%
        {hassan_et_al_2018}
\bibfield{author}{\bibinfo{person}{Mostafa Hassan}, \bibinfo{person}{Caterina
  Urban}, \bibinfo{person}{Marco Eilers}, {and} \bibinfo{person}{Peter
  M{\"u}ller}.} \bibinfo{year}{2018}\natexlab{}.
\newblock \showarticletitle{{MaxSMT}-Based Type Inference for {Python} 3}. In
  \bibinfo{booktitle}{\emph{Conference on Computer Aided Verification (CAV)}}.
  \bibinfo{pages}{12--19}.
\newblock
\urldef\tempurl%
\url{https://doi.org/10.1007/978-3-319-96142-2_2}
\showURL{%
\tempurl}


\bibitem[\protect\citeauthoryear{Hellendoorn, Bird, Barr, and
  Allamanis}{Hellendoorn et~al\mbox{.}}{2018}]%
        {deeptype}
\bibfield{author}{\bibinfo{person}{Vincent~J Hellendoorn},
  \bibinfo{person}{Christian Bird}, \bibinfo{person}{Earl~T Barr}, {and}
  \bibinfo{person}{Miltiadis Allamanis}.} \bibinfo{year}{2018}\natexlab{}.
\newblock \showarticletitle{Deep learning type inference}. In
  \bibinfo{booktitle}{\emph{Proceedings of the 2018 26th acm joint meeting on
  european software engineering conference and symposium on the foundations of
  software engineering}}. \bibinfo{pages}{152--162}.
\newblock


\bibitem[\protect\citeauthoryear{Maia, Moreira, and Reis}{Maia
  et~al\mbox{.}}{2011}]%
        {maia_moreira_reis_2011}
\bibfield{author}{\bibinfo{person}{Eva Maia}, \bibinfo{person}{Nelma Moreira},
  {and} \bibinfo{person}{Rog\'{e}rio Reis}.} \bibinfo{year}{2011}\natexlab{}.
\newblock \showarticletitle{A Static Type Inference for {Python}}. In
  \bibinfo{booktitle}{\emph{Workshop on Dynamic Languages and Applications
  (DYLA)}}.
\newblock
\urldef\tempurl%
\url{http://scg.unibe.ch/download/dyla/2011/dyla11_submission_3.pdf}
\showURL{%
\tempurl}


\bibitem[\protect\citeauthoryear{Malik, Patra, and Pradel}{Malik
  et~al\mbox{.}}{2019}]%
        {nl2type}
\bibfield{author}{\bibinfo{person}{Rabee~Sohail Malik}, \bibinfo{person}{Jibesh
  Patra}, {and} \bibinfo{person}{Michael Pradel}.}
  \bibinfo{year}{2019}\natexlab{}.
\newblock \showarticletitle{NL2Type: inferring JavaScript function types from
  natural language information}. In \bibinfo{booktitle}{\emph{2019 IEEE/ACM
  41st International Conference on Software Engineering (ICSE)}}. IEEE,
  \bibinfo{pages}{304--315}.
\newblock


\bibitem[\protect\citeauthoryear{Milner, Tofte, and Macqueen}{Milner
  et~al\mbox{.}}{1997}]%
        {10.5555/549659}
\bibfield{author}{\bibinfo{person}{Robin Milner}, \bibinfo{person}{Mads Tofte},
  {and} \bibinfo{person}{David Macqueen}.} \bibinfo{year}{1997}\natexlab{}.
\newblock \bibinfo{booktitle}{\emph{The Definition of Standard ML}}.
\newblock \bibinfo{publisher}{MIT Press}, \bibinfo{address}{Cambridge, MA,
  USA}.
\newblock
\showISBNx{0262631814}


\bibitem[\protect\citeauthoryear{Pradel, Gousios, Liu, and Chandra}{Pradel
  et~al\mbox{.}}{2020}]%
        {typewriter}
\bibfield{author}{\bibinfo{person}{Michael Pradel}, \bibinfo{person}{Georgios
  Gousios}, \bibinfo{person}{Jason Liu}, {and} \bibinfo{person}{Satish
  Chandra}.} \bibinfo{year}{2020}\natexlab{}.
\newblock \showarticletitle{Typewriter: Neural type prediction with
  search-based validation}. In \bibinfo{booktitle}{\emph{Proceedings of the
  28th ACM Joint Meeting on European Software Engineering Conference and
  Symposium on the Foundations of Software Engineering}}.
  \bibinfo{pages}{209--220}.
\newblock


\bibitem[\protect\citeauthoryear{Rak-amnouykit, McCrevan, Milanova, Hirzel, and
  Dolby}{Rak-amnouykit et~al\mbox{.}}{2020}]%
        {10.1145/3426422.3426981}
\bibfield{author}{\bibinfo{person}{Ingkarat Rak-amnouykit},
  \bibinfo{person}{Daniel McCrevan}, \bibinfo{person}{Ana Milanova},
  \bibinfo{person}{Martin Hirzel}, {and} \bibinfo{person}{Julian Dolby}.}
  \bibinfo{year}{2020}\natexlab{}.
\newblock \bibinfo{booktitle}{\emph{Python 3 Types in the Wild: A Tale of Two
  Type Systems}}.
\newblock \bibinfo{publisher}{Association for Computing Machinery},
  \bibinfo{address}{New York, NY, USA}, \bibinfo{pages}{57–70}.
\newblock
\showISBNx{9781450381758}
\urldef\tempurl%
\url{https://doi.org/10.1145/3426422.3426981}
\showURL{%
\tempurl}


\bibitem[\protect\citeauthoryear{Sulzmann and Hudak}{Sulzmann and
  Hudak}{2000}]%
        {Sulzmann00}
\bibfield{author}{\bibinfo{person}{Martin~Franz Sulzmann} {and}
  \bibinfo{person}{Paul Hudak}.} \bibinfo{year}{2000}\natexlab{}.
\newblock \emph{\bibinfo{title}{A General Framework for Hindley/Milner Type
  Systems with Constraints}}.
\newblock \bibinfo{thesistype}{Ph.D. Dissertation}. \bibinfo{address}{USA}.
\newblock
\showISBNx{0599791896}
\newblock
\shownote{AAI9973781.}


\bibitem[\protect\citeauthoryear{van Rossum, Lehtosalo, and Langa}{van Rossum
  et~al\mbox{.}}{[n.d.]}]%
        {PEP484}
\bibfield{author}{\bibinfo{person}{G. van Rossum}, \bibinfo{person}{J.
  Lehtosalo}, {and} \bibinfo{person}{L. Langa}.}
  \bibinfo{year}{[n.d.]}\natexlab{}.
\newblock \bibinfo{title}{{PEP484: Type Hints}}.
\newblock
  \bibinfo{howpublished}{\url{https://www.python.org/dev/peps/pep-0484/}}.
\newblock
\newblock
\shownote{[Online; accessed 8-February-2021].}


\bibitem[\protect\citeauthoryear{Xu, Zhang, Chen, Pei, and Xu}{Xu
  et~al\mbox{.}}{2016}]%
        {xu_et_al_2016}
\bibfield{author}{\bibinfo{person}{Zhaogui Xu}, \bibinfo{person}{Xiangyu
  Zhang}, \bibinfo{person}{Lin Chen}, \bibinfo{person}{Kexin Pei}, {and}
  \bibinfo{person}{Baowen Xu}.} \bibinfo{year}{2016}\natexlab{}.
\newblock \showarticletitle{Python Probabilistic Type Inference with Natural
  Language Support}. In \bibinfo{booktitle}{\emph{Foundations of Software
  Engineering (FSE)}}. \bibinfo{pages}{607--618}.
\newblock
\urldef\tempurl%
\url{http://doi.acm.org/10.1145/2950290.2950343}
\showURL{%
\tempurl}


\end{thebibliography}










\end{document}